\documentstyle[aps,epsf,preprint,prl]{revtex}

\def\beq{\begin{equation}}
\def\eeq{\end{equation}}
\def\beqn{\begin{eqnarray}}
\def\eeqn{\end{eqnarray}}
\def\biz {\begin{itemize}}
\def\eiz {\end{itemize}}
\def\ed{\end{document}}

\def\ed{\end{document}}

\begin{document}
%\draft

\title{Tunneling lifetime of trapped condensates}

\author{Nimrod Moiseyev$^{(1)}$ and Lorenz S. Cederbaum$^{(2)}$}

\address{{$^{(1)}$ Department of Chemistry and Minerva Center of
Nonlinear Physics in Complex Systems Technion -- Israel Institute
of Technology Haifa 32000, Israel.  $^{(2)}$Theoretische Chemie,
Physikalisch-Chemisches Institut, Universit\"{a}t Heidelberg, Im
Neuenheimer Feld 229, D-69120 Heidelberg, Germany.\\}}

\maketitle

\begin{abstract}
Using the complex scaling transformation  we calculate the
tunneling lifetime of a condensate inside a magnetic/optical trap.
 We show
that by varying the scattering length, the external potential acts
like a 'selective membrane' which controls the direction of the
flux of the cold atoms through the barriers and thereby controls
the size of the stable condensate inside the trap.
\end{abstract}

 Tunneling of Bose-Einstein condensates (BEC) through a dynamical potential barrier
has been demonstrated in the experiments of Phillips'
group\cite{phillips}. Recently it has been shown that the
tunneling of cold atoms can be controlled from full suppression to
strong enhancement by varying slightly the experimental
parameters\cite{va-so-nm}.  In contrast to the previous studies
where the tunneling is through a dynamical potential barrier,   we
discuss here the tunneling of the BEC from a potential well to the
continuum through a static potential barrier. In all BEC
experiments the condensates are trapped inside an open external
potential well which is embedded in between potential barriers.
Therefore, in the case of a positive scattering length (repulsive
atom-atom interactions) a given external open potential well
cannot trap a condensate  with more than a finite critical number
of atoms. As the number of the atoms in the condensate, N, is
increased, fractions of the condensate can tunnel through the
potential barriers.

 What is the lifetime of a trapped condensate as  N is
increased ? Can we control the number of atoms in the condensate
by varying slightly the scattering length ?
 Using the complex scaling transformation
which is based on rigorous mathematical
ground\cite{balslev-combes,simon} and  has been used before for
calculating the lifetimes of metastable states in atomic,
molecular and nuclear physics\cite{review-Reinhardt,review-NM} we
show that the lifetime of the condensates can be calculated and
the answer to the other question is positive as well.

% That
%is, by letting  a fraction of the condensate to tunnel through the
%potential barriers out of the trap, the reduced sized condensate
%is stabilized inside the external potential well "forever".
%However, as it will be demonstrated below, when the repulsive
%scattering length is sufficiently small it is possible to increase
%the number of the atoms in the trapped condensate by letting cold
%atoms to tunnel through the potential barriers from a reservoir of
%cold atoms which is located somewhere out of the trap into the
%external potential well in the trap. In this sense the external
%potential acts like a selective controlled membrane.

For the sake of clarity and without loss of generality let us
represent a simple one-dimensional external potential, $V_{ ext}
(x)=(x^2/2-0.8)\exp(-0.1x^2)$.  For a single atom this external
potential supports only a single bound state. Consequently, in the
absence of interaction between the atoms of the BEC, the
condensate can be found permanently trapped inside the potential
well or be  in one of the metastable states (so called resonances)
where the atoms can tunnel out of the trap (see inset in Fig.~1).
We shall see that in the presence of repulsive interaction between
the atoms the situation changes. Whether the condensate can be in
a bound state at all depends on the interaction strength and on
the number of atoms.

Before discussing our results we discuss in the following
metastable states and how they are determined by complex scaling
in the case of condensates.
%  The lifetime of the
%metastable states are equal to $\hbar/\Gamma$ where $\Gamma$ are
%the resonance widths.
 In hermitian quantum mechanics  metastable states are
associated with propagation of wavepackets escaping the potential
well.  However, in quantum scattering theory it is possible to
associate metastable states with the solutions of the
time-independent Schr\"odinger equation. The metastable states, as
are the bound states, are associated with outgoing boundary
conditions. In the case of BEC the metastable states are embedded
in the continuum part of the external potential. It is known from
scattering theory that outgoing solutions which are embedded in
the continuum are obtained only when the wave vector, k, gets
complex values $k=|k|\exp(-i\alpha)$, and hence the eigenvalues of
the Hamiltonian, $\{{\cal E}\}$, also become complex
quantities\cite{taylor}. Complex eigenvalues are obtained since
the corresponding eigenfunctions
 diverge exponentially, i.e., $\Psi_{res}\to \exp(\pm
i|k|\cos(\alpha)x)\exp(\pm|k|\sin(\alpha)x)\to +\infty$ as $x\to
\pm \infty$, and are not embedded in the hermitian domain of the
Hamiltonian.
%The real part of the complex energies, $\{E\}$ is
$\{Re{\cal E}\}$ are associated with the energies of the
metastable system and
%the imaginary parts, $\{\Gamma\}$,
$\{-2Im {\cal E}\}$, are associated with the rates of the decay
(inverse lifetimes). By complex scaling of the internal
coordinates in the Hamiltonian (in our case $x\to x\exp(i\theta)$
where $\theta \ge \alpha$)  the metastable states are taken back
into the Hilbert space and become square integrable (like the
bound states). This approach enabled  developing the quantum
theory for non-hermitian Hamiltonians (see the Reviews in
 Ref.\cite{review-Reinhardt}) and Ref.\cite{review-NM}.

How to compute the resonances of BEC ? The BEC  consists of N
atoms that are assumed to have contact potential interactions,
$U_0\delta(\vec{r}_j-\vec{r}_{j'})$, where $U_0= 4\pi
a_0\hbar^2/M$ and $a_0$ is the  s-wave scattering length\cite{GP}.
To calculate the bound and the resonance (metastable) states, as
explained above, we carry out the following similarity
transformation\cite{review-NM}, ${\cal{H}}_{BEC}=
\hat{S}\hat{H}_{BEC}\hat{S}^{-1}$, where $\hat{S}$ is the complex
scaling operator. The result of the complex scaling transformation
is, $ {\cal{H}}_{BEC}(\theta)=-{\hbar^2}/{2M}\sum_j
\left(\exp({-2i\theta})\nabla^2_j +V_{
ext}(\exp({i\theta})\vec{r}_j)\right) + U_0\exp({-i\theta})
\sum_{j,j'>j} \delta(\vec{r}_j-\vec{r}_{j'}).$ The bound and the
resonance states are obtained by solving the complex eigenvalue
problem, ${\cal{H}}_{BEC}(\theta)\Psi(\theta)={\cal E}
\Psi(\theta)$, where $\Psi(\theta)$ are square-integrable
functions and ${\cal E}$ are $\theta$-{\it independent}
eigenvalues. The bound states (i.e., $Im {\cal E}=0$) are obtained
in our case for any value of $0 \le \theta <\pi/4$. The resonance
energies and widths per atom, ${\cal E}/N\equiv E-i/2\Gamma$,  are
obtained when $\pi/4 > \theta
> \theta_c$. The value of the critical scaling angle is given by
$ 2\theta_c=\arctan{2E/\Gamma}$\cite{review-Reinhardt,review-NM}.
 Let us return to the statement that
$ \Psi(\theta)$ are square-integrable functions. As we will see
later this is an important point in the derivation of the
non-linear Schr\"odinger equation for complex scaled Hamiltonians.
Square integrability implies that the corresponding functions
decay exponentially to zero. But this is not enough. We need to
define the inner product. In hermitian quantum  mechanics where $
{\hat{H}}^{\dag}={\hat{H}}$ we use the common scalar product.
 In the case of
complex scaled Hamiltonians $ {\hat{H}}^{\dag}\ne{\hat{H}}$. In
our case (as in most studied physical cases for complex scaled
Hamiltonians\cite{review-NM} ), $
 {\cal{H}}_{BEC}^\dag(\theta)={\cal{H}}_{BEC}^*(\theta)$, and
therefore the c-product rather than the scalar product should be
used\cite{review-NM} such that, $
\left(\Psi_{j'}(\theta)|\Psi_j(\theta)\right)\equiv
\left<\Psi^*_{j'}(\theta)|\Psi_j(\theta)\right>=\delta_{j',j}$,
where $\Psi_{j'}(\theta)$  and $\Psi_j(\theta)$ are any bound or
resonance eigenfunctions of ${\cal{H}}_{BEC}(\theta)$ as defined
above. Note that here we assume that for $\theta=0$ (i.e.,
hermitian quantum mechanics) the eigenfunctions of ${\hat H}$ are
real.

%{\it Resonance energies and lifetimes associated with the
%non-linear Schr\"odinger equation}-
The non-linear Schr\"odinger equation, also known as the
Gross-Pitiavesky equation, is commonly applied to
condensates\cite{GP}. It is a mean field approximation to the full
many-body Hamiltonian of the problem as the Hartree-Fock
approximation is for fermions. All atoms are assumed to occupy the
same orbital and the total wavefunction is a product of these
orbitals\cite{GP}. Therefore, ${\cal
E}=<\Psi^*(\theta)|{\cal{H}}_{BEC}|\Psi(\theta)>$, where
$\Psi(\theta)=\prod_{j=1}^N \phi_{\theta}(\vec{r}_j)$. Minimizing
${\cal E}$  with respect to the orbital, keeping in mind the
definition of the inner product, leads to the complex scaled
version of the non-linear Schr\'odinger equation:
\begin{equation}
\left({\hat H}_0(\theta) + U
\exp({-i\theta})\phi_{\theta}^2\right)\phi_{\theta}=
(\mu-\frac{i}{2}\gamma) \phi_{\theta},
 \label{NLSE}
\end{equation}
where, ${\hat
H}_0(\theta)\equiv-{\hbar^2}\exp({-2i\theta})/{2M}\nabla^2 +V_{
ext}(\exp({i\theta})\vec{r} )$ and  $ U=U_0 (N-1) = {(4\pi a_0
\hbar^2})(N-1)/2M$ is the relevant non-linear parameter. The
complex chemical potential of the metastable state,
$\mu-\frac{i}{2}\gamma$ ($\gamma=0$ for bound states), is
associated with the complex energy of the BEC per atom, ${\cal
E}/N=E-i/2\Gamma$, as given by,
\begin{equation}
E-\frac{i}{2}\Gamma= \mu-\frac{i}{2}\gamma +
\frac{U}{2}\exp({-i\theta})\int_{all-space}\phi_{\theta}^4
d\vec{r}.
 \label{complexE}
\end{equation}
% As it will be shown below the bound and the resonance complex
% chemical potentials and complex energies that were obtained in
% our calculations are $\theta$ independent (for $0.7 \ge \theta\ge
% 0.3\, rad$).

 Within the framework of the
Gross-Pitiavesky approximation resonances can only be calculated
when the complex scaling transformation is applied  to the BEC
N-body Hamiltonian {\it before}  applying the mean field
approximation. To see why one cannot apply first the mean-field
approximation and only later the complex scaling transformation,
we start from the usual Gross-Pitiavesky Hamiltonian and scale it.
 The result is given by, $
 {\cal H}_{GP}(\theta)= \hat{S}\hat{\hat H}_{GP}\hat{S}^{-1}=
-\exp({-2i\theta}){\hbar^2}/{2M}\nabla^2 +V_{
ext}(\exp({-i\theta})\vec{r} ) + U
[\hat{S}\phi_{res}^*][\hat{S}\phi_{res}]. $ The resonance solution
of Eq.~\ref{NLSE} is square integrable. It is, however, associated
with the exponentially diverging unscaled resonance solution,
$\phi_{res}$, such that,
 $\phi_{\theta}=\hat{S}\phi_{res} \to
\exp(+i|k|\exp(+i(\theta-\alpha)) r)$  where $\theta > \alpha$.
Therefore, $[\hat{S}\phi_{res}^*][\hat{S}\phi_{res}]=
\exp(-i2|k|\sin(\theta)\sin(\alpha)
r)\exp(+2|k|\cos(\theta)\sin(\alpha) r) \to +\infty$ as $r\to
\infty$. Namely, the complex scaled non-linear potential term
$U[\hat{S}\phi_{res}^*][\hat{S}\phi_{res]}$ appearing in ${\cal
H}_{GP}(\theta)$ diverges as $r\to \infty$ (although
$\phi_{\theta}$ decays exponentially to zero) and its meaning is
unclear. For $U>0$ all states become bound and for $U<0$ the
potential is unbound from below. {\it Therefore, it is impossible
 by  this approach to get an eigenfunction which is
associated with the complex scaled resonance eigenfunction}. Note
that this difficulty has been avoided in a previous study of
bound-resonance state transitions using the Gross-Pitaevski
equation by inserting complex absorbing potentials in the regions
where it has been assumed that the atoms do not interact with one
another\cite{nm-band}. However, here we do not use this
approximation and the atoms can interact everywhere and not only
inside the external potential well.
%The question of how resonances in BEC
%should be calculated using such complex potentials beyond that
%assumption remains open and not relevant to the present work.

 We solved Eq.~\ref{NLSE} for the 1D potential defined
above, by increasing adiabatically the non-linear parameter U. We
used 400 particle-in-a-box basis functions as a basis set with the
box size of $L=50\,a.u.$. In each step of the calculations we
carried out self-consistent field iterative calculations to get
converged results in 8  significant figures where the scaling
angle $\theta$ has been varied from $0.3 rad$ to $0.7 rad$.  By
solving Eq.~\ref{NLSE} we calculated the complex
chemical-potential, $\mu-i/2\gamma$ and by solving
Eq.~\ref{complexE} we computated the complex-energy  per atom
 ${\cal E}/N=E-i/2\Gamma$.

 In Fig.~1 we show $\Gamma$ and $\gamma$ as functions of
$U$. We mark by arrow the critical value $U_c=0.8279$ for which
the bound-resonance transition occurs.
 For $U<U_c$ the system is bound, i.e. trapped
forever in the well, and for $U>U_c$ it is metastable and
tunneling through the barriers occurs. The trap potential is
depicted in the inset of the figure. The corresponding real parts
$\mu(U)$ of the chemical potential and E(U) of the energy are
shown in Fig.~2 as functions of U. As one can see, at the critical
value Uc where the resonances are "born" the chemical potential
vanishes, $\mu_c = \mu(U_c) = 0$. This is an expected result. We
may interpret $\mu$ as the energy needed to take a single atom of
the condensate out of the potential well. $\gamma$ is the
corresponding rate of decay, i.e. $1/\gamma$ is the tunneling time
of a single atom with a chemical potential $\mu$. The lifetime of
the BEC resonance state is $1/(N\Gamma)$, which can be short if
the condensate possesses many atoms. In a condensate all atoms are
equivalent and each of them can tunnel with the same probability.
Interestingly, $\Gamma>\gamma$ for all values of the non-linear
coupling $U >U_c$ as can be seen in Fig.~1. The reason is
persumably that $1/\Gamma$ is the lifetime per particle, and each
of the BEC's particles can tunnel through the barrier in different
ways, as a single particle or together with several other
equivalent particles. As mentioned above, $1/\gamma$, on the other
hand, is associated with the tunneling of a single atom only.

How to visualize the decay of a many-body system like a BEC? Once
the BEC is in a metastable state, a fraction of the atoms
consisting of (1-X)N atoms tunnels out of the potential well into
the continuum. This leads to a stabilization of the system where
XN atoms remain in the trap. Since all the atoms in the trap repel
each other, the energy of the XN atoms is lower than that of the N
atoms. Tunneling proceeds until the fraction of atoms which
remains in the trap forms a bound state. The number of atoms which
will tunnel clearly depends on the energy of the condensate.
However, within the framework of the Gross-Pitaevski mean-field
approximation the non-linear parameter is $U=U_0(N-1)$ where N is
the total number of atoms, and one cannot tell at a given value of
U which fraction of N will tunnel. As we will show below we can
enlighten this problem by taking into consideration the idea that
not always the best mean-field for condensates is obtained for the
case where all identical bosons of a condensate reside in a {\it
single} orbital\cite{lenz_BEC}.

We consider here the scenario where the ground state $\Psi$ is a
product of two types of spatial orbitals. There are $n_1$ atoms
which occupy the $\phi$ orbital and $n_2 = N-n_1$ atoms occupy the
$\chi$ orbital.
%Since the bosons are identical, the product is
%symmetrized (via the symmetrizer ${\cal S}$) and the wavefunction
%reads\cite{lenz_BEC}: $
%\Psi(\vec{r}_1,\vec{r}_2,...,\vec{r}_N)={\cal {S}}
%\phi(\vec{r}_1)\phi(\vec{r}_2)...\phi(\vec{r}_{n_1})
%\chi(\vec{r}_{n_1+1})\chi(\vec{r}_{n_1+2})...\chi(\vec{r}_N)$.
 We
associate the $n_1$ atoms in the orbital $\phi$ with the fraction
of the condensate $X= n_1/N$ which are located in the well and the
$n_2$ atoms in the orbital $\chi$ with the fraction $(1-X) =n_2/N$
which has tunneled through the barriers into the continuum.
Consequently, $\phi$ is located in the trap and $\chi$ well
outside the trap where $V_{ext} = 0$, and we may consider these
two orbitals to not overlap. Since $U=U_0(N-1)\sim U_0N$, the
energy per atom in the trap $E_{\phi}(X,U)$ is nothing but $E(XU)$
which is reported in Fig.~2. Analogously, the energy per atom in
the continuum is given by $E_{\chi}((1-X),U)$. The resulting
energy per atom of the whole system at a given U and X is thus $
E_{BEC}(X,U) = XE(XU) + (1-X)E_{\chi}((1-X),U)$. Because our
condensate is repulsive, the energy $E_{\chi}$ is known to be
equal to zero\cite{GP}  and we obtain the final result,
\begin{equation}
  E_{BEC}(X,U) = XE(XU),
 \label{EBEC}
\end{equation}
which makes clear that $E_{BEC}(X,U)$ can be derived from the
curve E(U) in Fig~2.  In complete analogy the rate of decay into
the continuum per atom corresponding to $E_{BEC}(X,U)$ is given by
$X\Gamma(XU)$ and is also already computed, see Fig.~1.

While for a given value of U the Gross-Pitiavesky energy E(U) is
just a number and does not provide us with the knowledge on how
many atoms have tunneled, $E_{BEC}(X,U)$ is the key to this
information. In Fig.~3 we show the energy per atom of the
condensate  as a function of X for different values of U. The
curves at the bottom of the figure are for small values of U,
those at the top for larger values of U. Each of the curves
exhibits a minimum at $X_c(U)$ and these minima play a central
role in the understanding of the tunneling process. These minima
are marked by solid dots. Let us consider a single curve in Fig.~3
for which $X_c$ is smaller than 1 (the value X=1 is marked in the
figure by a vertical line). X=1 implies that we have put N atoms
in the condensate and is thus our starting point. Because of the
variational principle, the condensate will minimize its energy by
letting a fraction $1-X_c$ of its atoms tunnel into the continuum
and keep the fraction $X_c$ in the well at which $E_{BEC}$ takes
on its minimum.

How does this appealing picture relate to the decay rate
$X\Gamma(XU)$ ? We know already from the scaled mean-field
approach that $\Gamma(U)$ changes from being zero to non-zero at
the bound-resonance transition point $U_c$. Therefore, for a given
value of U we find a particular value of X such that $XU=U_c$, and
this X tells us at which fraction of the condensate the system is
just still bound. If this is the case, there should be an intimate
relation between this particular value of X and $X_c$ at which
$E_{BEC}$ has its minimum.  {\it Indeed, we find that both values
of X are identical}, i.e. $X_c=U_c/U$. Using the value
$U_c=0.8279$ found above, one readily reproduces the values of X
at which any curve $E_{BEC}(X,U)$  takes on its minimum for a
given U. In particular, $E_{BEC}(X,U_c)$ -which is the blue curve
in Fig.~3- exhibits its minimum at $X_c=1$. This value implies
that for $U_c$ the condensate with N atoms is bound, while for
$U>U_c$ the fraction $1-X_c$ tunnels to make the remaining
fraction $X_c<1$ bound.

It is essential to note that the critical values $X_c$ {\it and}
$U_c$ can be determined from the above analysis without using the
complex scaled mean-field results. The curves $E_{BEC}(X,U)$ shown
in Fig.~3 can be computed via Eq.\ref{EBEC} for all values of U
from X=0 up to $X_c$ {\it using bound state calculations only}.
This is a success of the two-orbital picture\cite{lenz_BEC} used
above to derive Eq.~\ref{EBEC}.

Once $U \ge U_c$, increasing the s-wave scattering length $a_0$,
for instance,  by applying an external magnetic field to adjust
the relative energy of different internal states of the
atoms\cite{Regal}, leads to an increase of U and hence to a
reduction of the number of atoms inside the potential well (see
Fig.~3). On the other hand, if we decrease the scattering length
$a_0$, the trap can accommodate more atoms. Consequently, if there
is a reservoir of cold atoms outside the trap, {\it some of them
can tunnel through the barriers into  the trap} thus increasing
the number of atoms inside the trap.  Note that in Fig.~3 the
minima of $E_{BEC}$ for $U<U_c$ are at $X_c>1$, i.e, the
condensate inside the trap is further stabilized if atoms are
added. {\it In this way the size of the trapped condensate can be
controlled and the trap acts as a "controllable membrane" by
varying the s-wave scattering length (or by varying the depth of
the trap potential)}.  We hope that these fascinating results will
stimulate new experiments.

\begin{figure}[ht]
\epsfxsize=8.5cm \epsffile{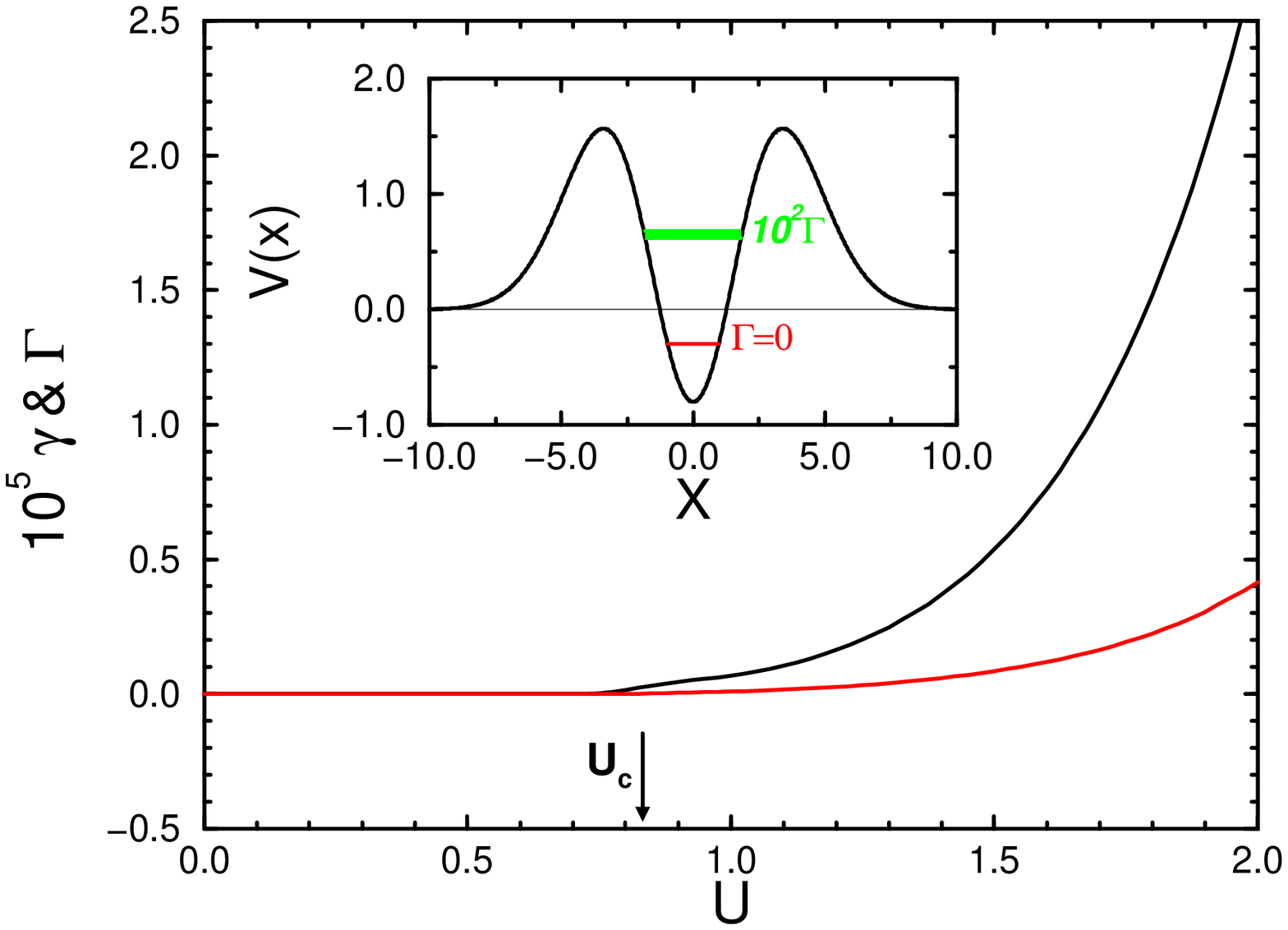} \caption
[kdv]{\label{Fig.1} {The rate of decay $\gamma$ of a single atom
and the total rate of decay per atom $\Gamma$ as a function of the
non-linear parameter U (see Eq.~\ref{NLSE} and text). The inset
shows the external trap potential used.}}
\end{figure}

\begin{figure}[ht]
\epsfxsize=8.5cm \epsffile{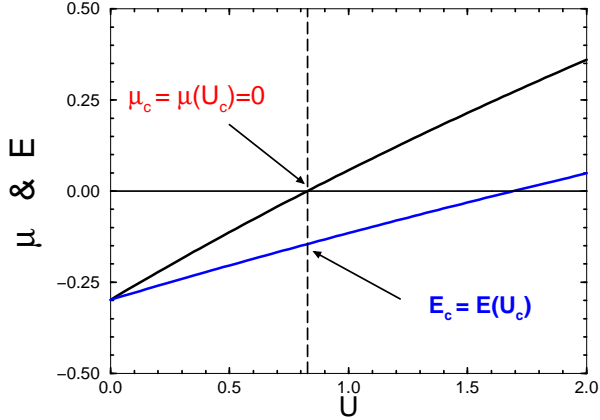} \caption [kdv]{\label{Fig.2}{
The chemical potential $\mu$(the real part of the complex
eigenvalue in Eq.~\ref{NLSE}) and the mean-field energy of the BEC
per atom E  (the real part of the complex energy ${\cal E}/N$, see
Eq.~\ref{complexE} in the text) as a function of the non-linear
parameter U.}}
\end{figure}

\begin{figure}[ht]
\epsfxsize=8.5cm \epsffile{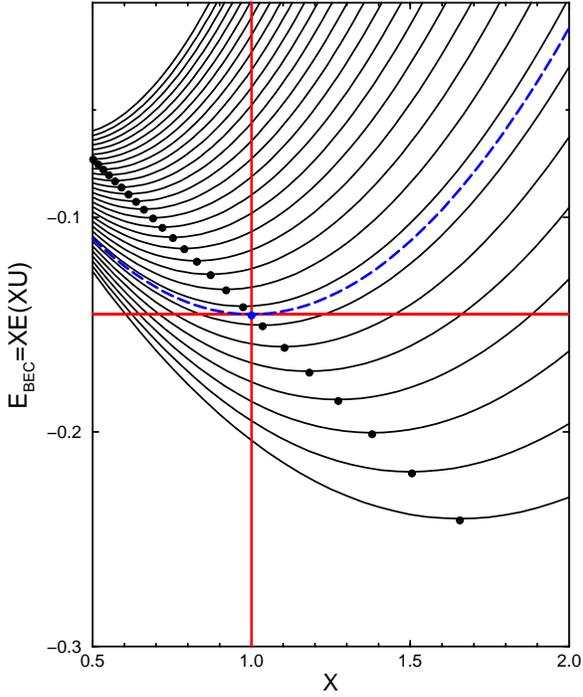} \caption [kdv]{\label{Fig.3}{
The energy per atom $E_{BEC}(X,U)$ as a function of the fraction
of atoms X which remains in the trap while a fraction 1-X of the
condensate has tunneled through the barriers into the continuum.
Each curve shown is for a different value of the non-linear
parameter U. From  bottom to top the j-th curve in black is
associated with U=0.5 + 0.05(j-1). The blue curve is for $U=U_c$.
The minima of the $E_{BEC}$ curves are at X=Xc (solid dots) and
play a central role in understanding the tunneling (see text).}}
\end{figure}

%\acknowledgments
% This work was
%supported in part by  the DFG and by the Fund for the Promotion of
%Research at the Technion .

\end{document}